\documentclass[11pt,a4paper]{article}
\pdfoutput=1
\usepackage{jheppub}

\usepackage{multirow, graphicx,amssymb,url,mathrsfs,amsmath}
\usepackage{wrapfig,boxedminipage,setspace,subfigure,epsfig}
\usepackage{amsxtra,amstext,latexsym,dsfont,amsfonts}
\usepackage{color,eucal}
\usepackage[dvipsnames]{xcolor}
\usepackage{float}
\usepackage{slashed,comment}
\usepackage{kotex}

\usepackage{tikz}
\usetikzlibrary{calc,patterns,angles,quotes}
\usetikzlibrary{decorations.pathreplacing,decorations.markings,snakes}

\newcommand{\kyr}[1]{{\color{red}{#1}}}

\title{Islands in charged linear dilaton black holes}

\author[a]{Byoungjoon Ahn,}
\author[a]{Sang-Eon Bak,}
\author[b,c]{Hyun-Sik Jeong,}
\author[a]{Keun-Young Kim,}
\author[b,c]{and Ya-Wen Sun}

\emailAdd{bjahn123@gist.ac.kr}
\emailAdd{sangeonbak@gm.gist.ac.kr}
\emailAdd{hyunsik@ucas.ac.cn}
\emailAdd{fortoe@gist.ac.kr}
\emailAdd{yawen.sun@ucas.ac.cn}

\affiliation[a]{School of Physics and Chemistry, Gwangju Institute of Science and Technology, \\
123 Cheomdan-gwagiro, Gwangju 61005, Korea}
\affiliation[b]{School of physics $\&$ CAS Center for Excellence in Topological Quantum Computation, University of Chinese Academy of Sciences, Zhongguancun east road 80, Beijing 100049, China}
\affiliation[c]{Kavli Institute for Theoretical Sciences, University of Chinese Academy of Sciences, \\ Zhongguancun east road 80, Beijing 100049, China}

\abstract{
We investigate the Page curve for a non-standard black hole  which is asymptotically non-flat/AdS/dS.
For this purpose, we apply the island prescription to the charged linear dilaton black holes and analyze, in detail, the entanglement entropy of Hawking radiation for both the non-extremal case and the extremal case. 
In the non-extremal case, we find the Page curve consistent with the unitarity principle: at early times the entanglement entropy grows linearly in time without the island and at late times it is saturated double of the Bekenstein-Hawking entropy in the presence of the island. We observe the Page time is universal for all different models studied by our method: $t_{\text{Page}}
    = \frac{3}{ \pi c}\frac{S_{\text{BH}}}{T_H}.$
For the extremal case, the island prescription provides the well defined entanglement entropy only with the island, which can not be obtained from the continuous limit of the non-extremal case. This implies that the Page curve may not be reproduced for the extremal case and further investigation is needed.
}

\begin{document}
\maketitle

\section{Introduction}
The black hole information is one of the fundamental problems in many areas of physics such as quantum mechanics, thermodynamics and the theory of general relativity~\cite{Hawking:1976aa,Hawking_1975,Page:1993wv,Wald_1975,Parker:1975aa}. 
One of the well-known black hole information issues is initiated from the relation between the entanglement entropy and the Hawking radiation. In 1975, Hawking proposed that the evaporating black hole undergoes the thermal process so that the black hole behaves as the thermal radiation: the Hawking radiation~\cite{Hawking_1975}. This implies that, as the black hole is evaporating from the pure state, the entanglement entropy outside the black hole is supposed to be increasing. However, this result is contrary to what the basic assumption of the quantum mechanics, the unitarity principle, requires: the entanglement entropy has to be zero at the end of the evaporation process since the final state still must be the pure state.

\paragraph{Information paradox and the Page curve:}
In addition to the evaporating black hole~\cite{Almheiri:2019hni,Almheiri:2019yqk,Almheiri:2019qdq,Almheiri:2019psf,Penington:2019npb,Engelhardt:2014gca}, the eternal black hole also has the similar information issue on the entanglement entropy. At the ``end stage'' of the evaporation, the evaporating black hole has the finite amount of the radiation so that the entanglement entropy is also bounded when the black hole vanishes. On the other hand, the eternal black hole has the infinite amount of the radiation, so does the entanglement entropy which is also contrary to the unitarity principle because unitarity requires the maximal limit of entropy of the black hole to be the Bekenstein-Hawking entropy~\cite{Bekenstein:1981aa}. 

The behavior of the entanglement entropy of the Hawking radiation is described by the Page curve~\cite{Page:1993wv,Page:2013dx,Page:1993df}. Thus, the information issue on the entanglement entropy of the Hawking radiation can be translated into how to reproduce the Page curve consistent with the unitarity principle, i.e. for the eternal black hole, the entanglement entropy is increasing and has to be bounded by the Bekenstein-Hawking entropy. 
Since resolving the information issue with the Page curve is related to rendering the gravity physics be coherent with the quantum mechanics, it is an important and essential to understanding quantum gravity.

\paragraph{Islands formula and the Page curve:}
In order to calculate the Page curve of Hawking radiation, it is recently proposed that the island formula~\cite{Engelhardt:2014gca,Akers:2019lzs,Faulkner:2013ana,Wall:2012uf,Almheiri:2019psf,Almheiri:2019hni} for the entanglement entropy of the Hawking radiation, $S(R)$, can deduce the behavior of the Page curve for a unitary evolution, which reads
%
%
%
\begin{align}\label{ISFOR}
\begin{split}
    S(R) = \text{min} \,\, \text{ext}\left[S_{\text{gen}} \right] \,, \qquad 
    S_{\text{gen}} = \frac{\text{Area}(\partial I)}{4 G_N}+S_{\text{matter}} (\text{R} \cup \text{Island})\,,
\end{split}
\end{align}
where $G_{N}$ is the Newton constant, $R$ is the radiation region.
$\partial I$ is the boundary of the island and $S_{\text{matter}}$ is the entropy of quantum fields.

Let us give the quick description of how this formula works. For more details, see \cite{Almheiri:2019hni,Page:1979tc} and references therein.
First, one can introduce the generalized entropy functional $S_{\text{gen}}$ containing two pieces as i) $\text{Area}(\partial I)/4 G_N$: the Bekenstein-Hawking entropy of the island; ii) $S_{\text{matter}} (\text{R} \cup \text{Island})$: the von Neumann entropy of the matter sector on the union of radiation ($R$) and the island region.
Next, one can evaluate $S_\text{gen}$ with respect to all possible saddle points (or extrema), $\text{ext}\left[S_{\text{gen}} \right]$, which is corresponding to the locations of the island and when its minimum value exists we can find $S(R)=\text{min} \,\, \text{ext}\left[S_{\text{gen}} \right]$.
In summary, the entanglement entropy of Hawking radiation $S(R)$ is identified with the generalized entropy $S_{\text{gen}}$ giving the minimum value over the choice of location of the islands.

\paragraph{Islands in general black hole in higher dimension:}
We further note two things of the island formula. 
First, AdS space (and holography) is not a necessary condition for the island formula. Although this island prescription is first suggested in the context of the holography (or AdS/CFT correspondence) by Ryu and Takayanagi~\cite{Ryu:2006bv} and its further developments~\cite{Hubeny:2007xt,Lewkowycz:2013nqa,Barrella:2013wja,Faulkner:2013ana,Engelhardt:2014gca,Penington:2019npb,Almheiri:2019psf,Almheiri:2019hni}, it can be applied to any quantum system coupled to gravity. This is further supported by the fact that the entanglement entropy has been found to follow the Page curve in all the examples (e.g., asymptotically flat or dS) studied in the literature so far with the island prescription.  Furthermore, using the gravitational replica method, it is shown that the island formula can be derived from the Euclidean path integral without holography~\cite{Almheiri:2019qdq,Penington:2019kki,Hartman:2020swn,Goto:2020wnk}. Thus holography or AdS space are not necessarily required to study the entropy of systems.

Second, the island formula can be applied to the higher dimensional spacetime~\cite{Penington:2019npb,Almheiri:2019psy,Hashimoto:2020cas,Karananas:2020fwx,Wang:2021woy,Yu:2021cgi,Alishahiha:2020qza,Geng:2020qvw}. The initial study of the Page curve was investigated for two-dimensional black holes using the semiclassical method in Jackiw-Teitelboim (JT) gravity~\cite{Almheiri:2019qdq,Almheiri:2019hni,almheiri2019islands} and the most of the research on the information issue are focused on the two-dimensional gravity systems where more tractable analysis is allowed. For the two-dimensional case, the island appears at the end stage of the evaporation and the Page curve is produced. It is argued that, in the higher dimensional systems, the island should appear and the Page curve in a unitary evolution also can be reproduced when the island is taken in accounted~\cite{Penington:2019npb}. Recently, the literatures supporting this argument are reported. For instance, the Schwarzschild black holes~\cite{Hashimoto:2020cas}, the Reissner-Nordstrom black holes~\cite{Wang:2021woy,kim2021entanglement}, and the dilaton black holes~\cite{Karananas:2020fwx,Yu:2021cgi} are considered in the higher dimensions. 
For the recent developments in this direction, see \cite{Almheiri:2020cfm,Penington:2019npb,Almheiri:2019psf,Almheiri:2019hni,almheiri2019islands,Chen:2019uhq,Almheiri:2019psy,Penington:2019kki,Almheiri:2019qdq,Chen:2019iro,Gautason:2020tmk,Anegawa:2020ezn,Hashimoto:2020cas,Karananas:2020fwx,Wang:2021woy,Hartman:2020swn,Hollowood:2020cou,Krishnan:2020oun,Alishahiha:2020qza,Banks:2020zrt,Geng:2020qvw,Chen:2020uac,Chandrasekaran:2020qtn,Li:2020ceg,Bak:2020enw,Bousso:2020kmy,Hollowood:2020kvk,Krishnan:2020fer,Engelhardt:2020qpv,Karlsson:2020uga,Gomez:2020yef,Chen:2020jvn,Hartman:2020khs,Balasubramanian:2020coy,Balasubramanian:2020xqf,Sybesma:2020fxg,Chen:2020hmv,Ling:2020laa,Marolf:2020rpm,Hernandez:2020nem,Matsuo:2020ypv,Goto:2020wnk,Akal:2020twv,KumarBasak:2020ams,Caceres:2020jcn,Raju:2020smc,Chu:2021gdb,Bhattacharya:2020ymw,Bhattacharya:2020uun,Manu:2020tty,Yu:2021cgi,kim2021entanglement,Wang:2021mqq,Caceres:2021fuw,Bhattacharya:2021jrn,Ghosh:2021axl,Geng:2020fxl,Geng:2021wcq,Geng:2021iyq,Geng:2021hlu} and the references therein.

\paragraph{Motivation of this paper:}
Although the island structure and the Page curve are investigated with the various black hole geometries in higher dimensions, to our knowledge, most of them considered the black holes in asymptotically flat/AdS/dS. 
Thus, it is worthwhile to verify that whether the island formula can be applied to other cases (non-asymptotically flat/AdS/dS) which is called the non-standard black hole geometries.
Checking the Page curve for the non-standard black hole geometries using the island formula is important not only for the range of applicability of the island method, but also for the quantum gravity.
In this paper, we make one step further in this direction.

For this purpose, we choose a charged linear dilaton black hole in four dimensions~\cite{Karananas:2020fwx}. This model is advantageous because the analytic background geometry solution is allowed. 
The main focus of~\cite{Karananas:2020fwx} is the case {\it without} the charge. We generalize the analysis in the presence of charge, for both the non-extremal and extremal cases.  The extremal case is addressed  in ~\cite{Karananas:2020fwx} but we find that we need to revisit the analysis for two reasons.
%
First, Ref.~\cite{Karananas:2020fwx} reported $S(R)$ {\it without} the island and the explicit computation of $S(R)$ {\it with} the island is not shown.
Second, the computation in \cite{Karananas:2020fwx} is based on the Penrose diagram of the {\it non-extremal} case, which turns out to be different from the extremal case.
It is shown that, in order to study the complete picture of the island, one needs to perform the separate analysis for the \textit{extremal} black hole and the \textit{non-extremal} black hole because they are essentially different~\cite{Carroll:2009maa}. For instance, the entanglement entropy, $S(R)$, in the extremal case can not be obtained from the continuous extremal limit of the non-extremal case~\cite{kim2021entanglement}.

This paper is organized as follows. 
In section 2, we introduce the charged linear dilaton black hole and discuss its properties.
In section 3, we review the method to calculate the entanglement entropy without and with the island.
In section 4, using the island formula, we study the entanglment entropy of Hawking radiation for the non-extremal black holes. 
In section 5, we analyze the entropy of the extremal black holes.
Section 6 is devoted to conclusions.

\paragraph{Note added:}
while we were finishing our project, we noticed a related paper \cite{Yu:2021cgi} which studies similar topics but in a  different model.

\section{The charged dilaton black hole} \label{section2}
Let us consider the four-dimensional dilaton action with a $U(1)$ gauge field in the Einstein frame 
\begin{equation}\label{ACTION}
    I=\frac{1}{16 \pi G_N} \int d^4 x \sqrt{g}\left(R-\frac{1}{2}\left(\partial \sigma\right)^2+4 k^2 e^\sigma -\frac{1}{4}e^{\gamma \sigma} F_{\mu \nu}F^{\mu \nu}\right)\,,
\end{equation}
where $k$ and $\gamma$ are constants, $\sigma$ is a scalar field, and $F_{\mu \nu}=\partial_\mu A_\nu - \partial_\nu A_\mu$ is a field strength tensor.
It is originated from the linear dilaton model in the string frame.
By varying the action with respect to $g_{\mu \nu}$, $A_\mu$, and $\sigma$, we have the equations of motion
\begin{equation}
    R_{\mu \nu}-\frac{1}{2} g_{\mu \nu} R =\frac{1}{2}\partial_\mu \sigma \partial_\nu \sigma - \frac{1}{2} g_{\mu \nu} \left(\frac{1}{2}\left(\partial \sigma\right)^2-4 k^2 e^\sigma \right)+\frac{1}{2} e^{\gamma \sigma}\left(F_{\mu \lambda} F_{\nu}^{\lambda} -\frac{1}{4} g_{\mu \nu} F_{\alpha \beta} F^{\alpha \beta} \right),
\end{equation}
\begin{equation}
    \nabla_\mu \left(e^{\gamma \sigma} F^{\mu \nu}\right) = 0\,,
\end{equation}
\begin{equation}
    \square \sigma = -4 k^2 e^\sigma +\frac{1}{4} \gamma e^{\gamma \sigma} F_{\mu \nu} F^{\mu \nu}\,.
\end{equation}
The equations of motion are satisfied by the following charged dilaton black hole solution
\begin{equation} \label{eq_first metric}
    ds^2=-r^2\left(1-\frac{2M}{r^2}+\frac{Q^2}{4 r^4}\right) dt^2 + \left(1-\frac{2M}{r^2}+\frac{Q^2}{4 r^4}\right)^{-1} dr^2 + r^2\left(dx^2+dy^2\right) \,,
\end{equation}
\begin{equation}
    A_t = -\frac{\mu}{2 \gamma} \left(kr\right)^{2\gamma}, \,\,\, A_r=A_x=A_y=0\,,
\end{equation}
\begin{equation}
    \sigma = -2 \log{\left(kr\right)}\,, 
\end{equation}
where $Q=\frac{\mu}{\sqrt{2}k}$ and $ \gamma=-1$. With respect to the metric (\ref{eq_first metric}), we consider two cases separately: (i) the \textit{extremal} black hole and (ii) the \textit{non-extremal} black hole.

\paragraph{(i) Non-extremal case}
In the case of $ 2M > Q $,  there exist an inner horizon $(r_{-})$ and an outer horizon $(r_{+})$ on the metric (\ref{eq_first metric}).

\begin{equation}
    r_{\pm} = \sqrt{M \pm \sqrt{M^2-\frac{Q^2}{4}}}\,.
\end{equation}

By using the analogy with the Reissner-Nordstr\"om black hole case, one can rescale $t$ into $r_+ t$. Then, by taking into account the null geodesic, one can introduce the tortoise coordinate:
\begin{equation}
\begin{split}
    r^* &= r_+ \int \frac{dr}{r\left(1-\frac{2M}{r^2}+\frac{Q^2}{4 r^4}\right)} \\
    &= \frac{1}{2\kappa_{+}}\left(\log{\frac{|r-r_{+}|}{r_{+}}}+\log{\frac{|r+r_{+}|}{r_{+}}}\right)+\frac{1}{2\kappa_{-}}\left(\log{\frac{|r-r_{-}|}{r_{-}}}+\log{\frac{|r+r_{-}|}{r_{-}}}\right)\,.
\end{split}
\end{equation}
Here, $\kappa_{\pm}=\frac{r_{\pm}^2-r_{\mp}^2}{r_+ r_{\pm}^2}$ are the surface gravity at each horizon. The line element becomes
\begin{equation}
    ds^2=\frac{(r^2-r_-^2)(r^2-r_+^2)}{r_+^2 r^2}\left(-dt^2+dr^{*2}\right) + r^2\left(dx^2+dy^2\right)\,.
\end{equation}

In terms of the Kruskal coordinate defined as
\begin{equation}
u = -e^{- \kappa_{+} u^*} =  -e^{-\kappa_{+}\left(t-r^*\right)}\,, \,\,\,\,\,\,  v = e^{\kappa_{+} v^*} = e^{\kappa_{+}\left(t+r^*\right)}\,,
\end{equation}
the line element reads
\begin{equation}
    ds^2 = - f(r)^2 du dv + r^2 \left(dx^2+dy^2\right)\,,
\end{equation}
where the conformal factor is given by
\begin{equation}
   f(r)^2 = \frac{r_{-}^2}{\kappa_{+}^2 r^2} \left(\frac{r_{-}^2}{|r-r_{-}||r+r_{-}|}\right)^{\frac{\kappa_{+}}{\kappa_{-}}-1}.
\end{equation}
Note that this coordinate is singular at the inner horizon $r=r_{-}$ and is regular in $(r_{-},\infty)$. One can not define a coordinate that is non-singular on both horizons simultaneously.

\paragraph{(ii) Extremal case}
For the case $Q = 2M$, two horizons coincide so that the metric of the extremal charged dilaton black hole (\ref{eq_first metric}) is written by
\begin{equation}
    ds^2 = -r ^2 \left(1-\frac{r_h^2}{r^2}\right)^2 dt^2 + \frac{dr^2}{ \left(1-\frac{r_h^2}{r^2}\right)^2} +r^2\left(dx^2 + dy^2\right)\,,
\end{equation}
and the event horizon locates at $r_h$. As $r\rightarrow \infty$ or $r_h\rightarrow 0$, the geometry approaches to 
\begin{equation}
    ds^2 = -r ^2  dt^2 + dr^2 +r^2\left(dx^2 + dy^2\right)\,.
\end{equation}
Note that a causal structure is not asymptotically flat and it has a naked singularity at $r=0$.

In the case of the extremal charged dilaton black hole, by transforming $t$ into $r_h t$  and taking into account the null geodesic, one can take the following tortoise coordinate.
\begin{equation}
    r^* = r_h \int \frac{dr}{r\left(1-\frac{r_h^2}{r^2}\right)^2}\,,
\end{equation}
in terms of which the metric is
\begin{equation}
    ds^2 = \frac{(r^2-r_h^2)^2}{r_h^2 r^2} \left(-dt^2+dr^{*2}\right)+r^2\left(dx^2 + dy^2\right)\,.
\end{equation}
By defining the Kruskal coordinates as
\begin{equation}
u = -e^{- \frac{u^*}{r_h}} =  -e^{-\frac{\left(t-r^*\right)}{r_h}}\,, \,\,\,\,\,\,  v = e^{ \frac{v^*}{r_h}} =  e^{\frac{\left(t+r^*\right)}{r_h}}\,,
\end{equation}
we can rewrite the metric 
\begin{equation}
    ds^2 = - f(r)^2 du dv + r^2 \left(dx^2+dy^2\right)\,,
\end{equation}
where the conformal factor is given by
\begin{equation}\label{ILLNEESNEEDSTOBECURE}
    f(r)^2 = \frac{(r^2-r_h^2)^2}{r^2} e^{-\frac{2r^{*}}{r_h}}\,.
\end{equation}
Although the conformal factor of the charged dilaton black hole is different from one of the Reissner-Nordstr\"om black hole, the causal structure of both are the same.
We will use the Kruskal coordinate to derive the entanglement entropy of the extremal charged dilaton black hole.

 \section{Review on the approach} \label{section3}

Our goal is to derive the Page curve from the entanglement entropy of the Hawking radiation of the charged linear dilaton black holes by calculating it with/without the islands.

One can identify the Hawking radiation of the black hole with a matter sector coupled to the gravitational theory. 
In two dimensional setup, this corresponds to a free CFT with $N \gg 1$ minimally coupled massless scalar fields, where the central charge c is $\mathcal{O}(N)$, but much smaller than a mass of the black hole.
The region of the Hawking radiation $R$ is represented by the union of $R_+$ and $R_-$ in the right and left wedge, respectively. We assume the distance between the event horizon and the reservoir are large enough to ignore the backreaction of the matter fields on the geometry.

In four dimensional models, the general expression of the entanglement entropy is not known. Fortunately, the charged dilaton black hole has a two-dimensional planar horizon so that we can deal with a density of the entanglement entropy on $\mathbb{R}^2$. Therefore, we apply the well-known results to our case, which is obtained in two-dimensional field theories.

We follow the way to calculate the entanglement entropy which is suggested in ~\cite{Hashimoto:2020cas}. Let us briefly review the argument.
One can consider two configurations, one of which has islands and the other does not.

For the case without islands, the finite part of the matter entanglement entropy comes from the separate two regions $R_+$ and $R_-$. Although the mutual information does not imply the entanglement itself, one can assume as a necessary condition that it is dealt with the finite part of the entanglement entropy between two regions.
Therefore, the finite part of the matter entanglement entropy is given by
\begin{equation}
   S(R)= -I(R_+; R_-)\,,
\end{equation}
where the mutual information is defined by
\begin{equation}
    I(A;B) \equiv -S(A \cup B) + S(A) +S(B)\,.
\end{equation}

In the limit the distance between the boundary surfaces is large, the mutual information is approximated by that of the two-dimensional massless fields.
Therefore, the entanglement entropy of the matter part without island is given by
\begin{equation}\label{eq_wo island}
    S_{\text{matter}}=\frac{c}{3} \log{\left[l(b_+,b_-)\right]}\,,
\end{equation}
where $b_+ = (t_b, b)$ and $b_- = (-t_b-i \pi / \kappa_+, b)$ indicate the cutoff surface for the left and right wedges. Here, the geodesic distance $l$ can be computed by
\begin{equation}
    l(z,z')=\sqrt{f(z)f(z')(u(z')-u(z))(v(z)-v(z'))}\,.
\end{equation}

In the configuration with the island, each of two boundaries of the island becomes much closer to the boundary of $R$ in the same wedge than to the boundaries in the other wedge at late times. In the right wedge, the dominant contribution comes from the finite part of the entanglement entropy between $R_+$ and the island.
The contribution from the left wedge is equal to the right one, so the matter entanglement entropy is twice of one in the right wedge.
Therefore, the finite part of the matter entanglement entropy is given by
\begin{equation}
   S(\text{R} \cup \text{Island})=-2I(R_+; \text{Island})\,.
\end{equation}
Assume that the cutoff surface is located far from the horizon, i.e. $r_+ \ll b$. Then, one can obtain the entanglement entropy of the two-dimensional massless fields which live on the radiation region $(R)$ and the island region
\begin{equation} \label{eq_w island}
    S_{\text{matter}}=\frac{c}{3} \log{\left[\frac{l(a_+,a_-)l(b_+,b_-)l(a_+,b_+)l(a_-,b_-)}{l(a_+,b_-)l(a_-,b_+)}\right]}\,.
\end{equation}
where $a_+ = (t_a, a)$ and $a_- = (-t_a-i \pi / \kappa_+, a)$ denote the boundary of the island.

\section{Entanglement entropy of non-extremal charged dilaton black holes}

\begin{figure}
\begin{center}
    \subfigure[Without island]
    {\includegraphics[width=7cm]{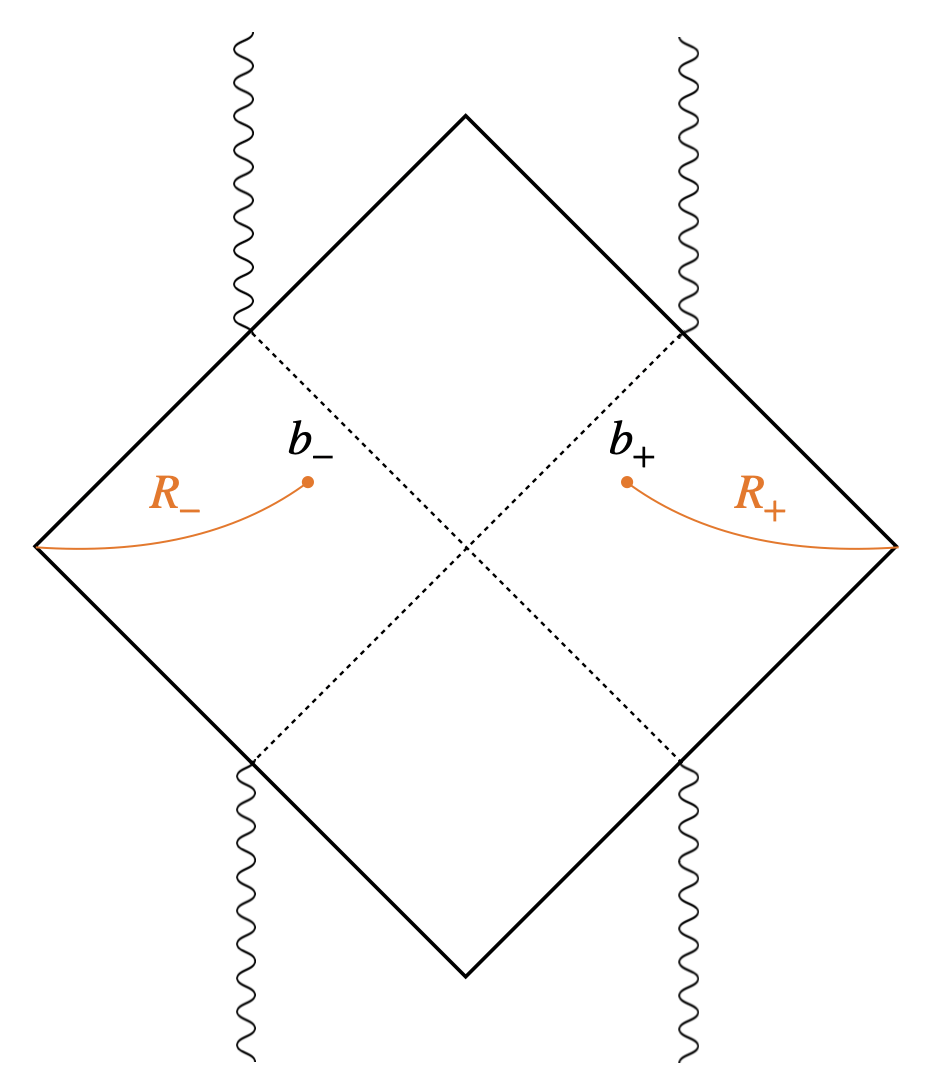}}
    \quad \quad
    \subfigure[With island] 
    {\includegraphics[width=7cm]{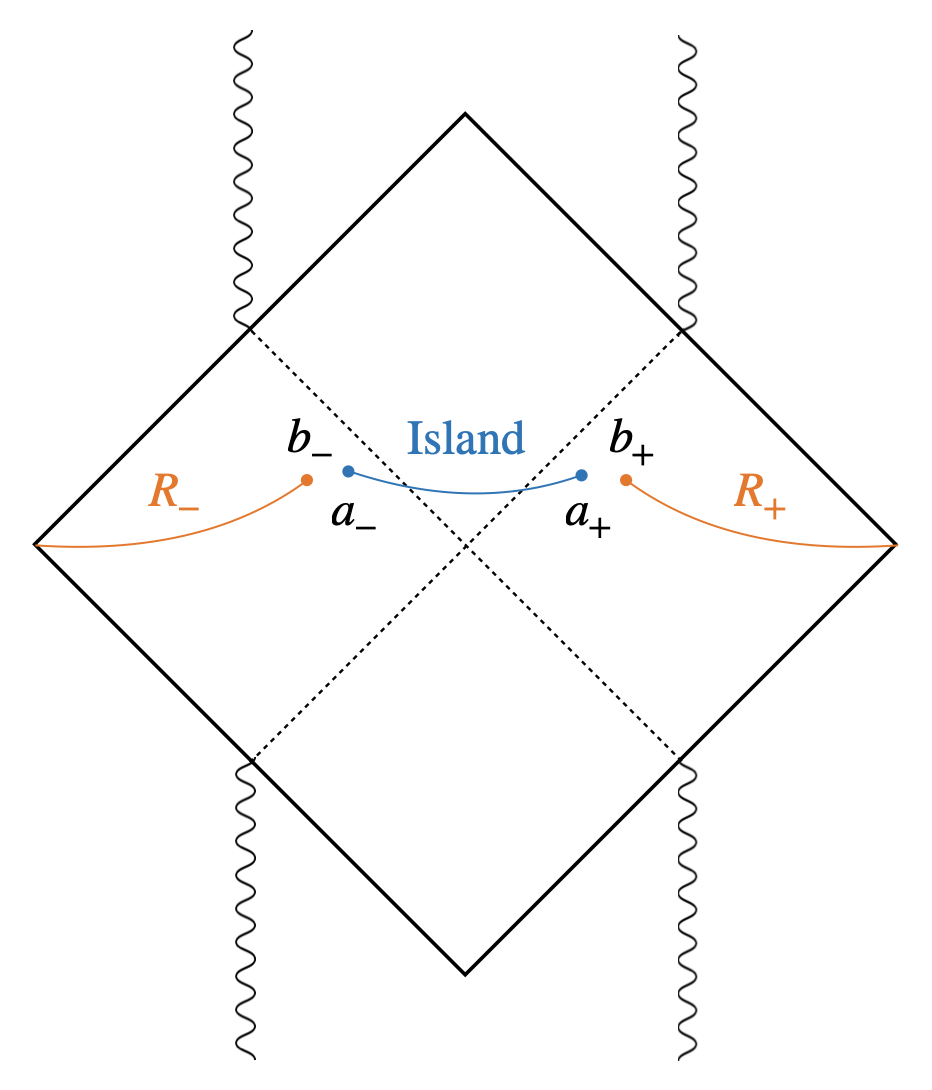}}
       
    \caption{(a) Penrose diagram of the non-extremal charged dilaton black holes without island. The union of $R_+$ and $R_-$ indicates the radiation region. The boundaries of $R_+$ and $R_-$ are denoted by $b_+$ and $b_-$, which indicate cutoff surfaces. (b) Penrose diagram of the non-extremal charged dilaton black holes with island. The boundaries of the island correspond to $a_+$ and $a_-$.}
\label{fig_non extremal}
\end{center}
\end{figure}

In this section, we derive the entanglement entropy with or without island configuration in the four-dimensional non-extremal charged dilaton black holes. Using these result, the Page curve is discussed in the context of black hole information paradox, with the Page time and scrambling time for our model. We employ the assumptions in section \ref{section3} to make the computation on the two-dimensional field theory sensible. It means, we will follow the approach that is firstly argued in \cite{Hashimoto:2020cas}.

\subsection{Entropy without island}

In this section, we consider the entropy of the matter sector in the absence of the island configuration. The entropy of the matter field is computed semiclassically on the charged dilaton black hole background.

By using the equation (\ref{eq_wo island}), one could write the  entropy of the matter fields without islands in terms of the conformal factor $f(b)$, surface gravity $\kappa_{\pm}$, and tortoise coordinate $r^*$
\begin{equation}
    S(R) = S_{\text{matter}} = \frac{c}{3} \log{\left[2 f(b) e^{\kappa_{+}r^*(b)}\cosh{\kappa_{+}t}\right]}\,.
\end{equation}

Assuming $r_+ \ll b$, we find that the entanglement entropy increases linearly in time
\begin{equation} \label{eq_entropy wo island}
    S_{\text{matter}} = \frac{c}{3} \log{\left(2 \cosh{\kappa_+ t} \right)} \simeq \frac{c}{3} \kappa_+ t\,.
\end{equation}
In the last approximation, we look into the late time behavior $t \gg b$, where $b$ is much larger than $r_+$. This linear growth is problematic because the entanglement entropy increases forever and exceed the entropy of the black hole in the end. Instead, we expect that the growth of the entanglement entropy will finish in a finite time. Finally, the entanglement entropy will saturate to twice the Bekenstein-Hawking entropy for the unitarity of the black holes. To resolve this problem, we will introduce the island configuration and check that this prescription provides the correct computation of the entanglement entropy for the charged dilaton gravity.

\subsection{Entropy with island}

In the presence of the island, one should take into account the island contribution when considering the generalized entropy. Consider the case that the observer outside the black hole collects the Hawking quanta which cross the cutoff surfaces $b_+ = (t_b, b)$ and $b_- = (-t_b-i \pi / \kappa_+, b)$ in Fig.~\ref{fig_non extremal}. It means that the degree of freedom of the radiation is counted with respect to region $R \equiv R_- \cup R_+$. However, the matter section of the generalized entropy contains the island region whose boundary is denoted by $a_+ = (t_a, a)$ and $a_- = (-t_a-i \pi / \kappa_+, a)$. We will see that the generalized entropy including an island contribution provides a correct description of the time evolution of the entropy of radiation.

By using the equation (\ref{eq_w island}), one could calculate the entanglement entropy of the matter field in the presence of island
\begin{equation}
\begin{split}
    S_{\text{matter}} =&
    \frac{c}{6} \log{\left[2^4 f^2(a) f^2(b) e^{2\kappa_+ \left(r^*(a)+r^*(b)\right)}\cosh^2{(\kappa_+ t_a)\cosh^2{(\kappa_+ t_b)}}\right]} \\
    &+ \frac{c}{3} \log{\left[\frac{\cosh{\kappa_+\left(r^*(a)-r^*(b)\right)}-\cosh{\kappa_+ (t_a-t_b)}}{\cosh{\kappa_+\left(r^*(a)-r^*(b)\right)} + \cosh{\kappa_+ (t_a+t_b)}}\right]}\,.
\end{split}
\end{equation}

Using this formula, we want to consider the early time and late time behaviors of the entropy. Before investigating them, one can impose that the cutoff surface is located far away from the outer horizon $r_+ \ll b$, obtaining the following expression
\begin{equation}
\begin{split}
    S_{\text{matter}} \simeq&\,\, \frac{c}{6}\log{\left[\frac{2^4 b^2}{\kappa_+^4 a^2} \left(\frac{|a^2-r_+^2|}{r_+^2}\right)\left(\frac{|a^2-r_-^2|}{r_+^2}\right) \cosh^2{(\kappa_+ t_a)} \cosh^2{(\kappa_+ t_b)}\right]} \\ &+\frac{c}{3}\log{\left[\frac{1-2\left(\frac{|a^2-r_+^2|}{b^2}\right)^\frac{1}{2} \left(\frac{|a^2-r_-^2|}{b^2}\right)^\frac{\kappa_+}{2\kappa_-}\cosh{\left(\kappa_+(t_a-t_b)\right)}}{1+2\left(\frac{|a^2-r_+^2|}{b^2}\right)^\frac{1}{2} \left(\frac{|a^2-r_-^2|}{b^2}\right)^\frac{\kappa_+}{2\kappa_-}\cosh{\left(\kappa_+(t_a+t_b)\right)}}\right]}\,.
\end{split}
\end{equation}

\paragraph{Early time}
The early time behavior of the entanglement entropy is obtained by taking into account the limit $t_a,\, t_b \ll r_+$. We can also assume that the extremal surface is located near the outer horizon.

The generalized entropy is written as the sum of the area term with respect to boundaries of the island and the entropy of the quantum matter (\ref{ISFOR})
\begin{align*}
    S_{\text{gen}} = \frac{\text{Area}(\partial I)}{4 G_N}+S_{\text{matter}} (\text{R} \cup \text{Island})\,.
\end{align*}

Note that the boundary of the island appears as planar geometry as we discussed in section \ref{section3}. Combining the above facts, the generalized entropy is written as
\begin{equation}\label{eq_with island early time}
\begin{split}
    S_{\text{gen}} \simeq& \,\,\frac{a^2}{2 G_N} + \frac{c}{6} \log{\left[\frac{2^4 b^2}{\kappa_+^4 a^2} \left(\frac{|a^2-r_+^2|}{r_+^2}\right)\left(\frac{|a^2-r_-^2|}{r_+^2}\right) \cosh^2{\left(\kappa_+ t_a\right)} \cosh^2{\left(\kappa_+ t_b\right)}\right]} \\ &-\frac{4c}{3}\left(\frac{|a^2-r_+^2|}{b^2}\right)^\frac{1}{2} \left(\frac{|a^2-r_-^2|}{b^2}\right)^\frac{\kappa_+}{2\kappa_-}  \cosh{\left(\kappa_+ t_a\right)} \cosh{\left(\kappa_+ t_b\right)}\,.
\end{split}
\end{equation}

To compute the entanglement entropy, one should find the position of $a$ extremizing (\ref{eq_with island early time}) over all possible Cauchy surfaces. Similar to the result in \cite{Wang:2021woy}, at early times one can find that $S_{\text{gen}}$ is extremized in the vicinity of the singularity

\begin{equation} \label{eq_plank island}
   a \simeq \sqrt{\frac{c}{3}} l_P\,,
\end{equation}
where $l_P $ is the Plank length. It seems that there is an island at a distance of Plank length from the singularity. However, this extremal point (\ref{eq_plank island}) can not be the boundary of the island. This is because we are taking into account the Cauchy surface which only covers outside the inner horizon (see Fig.~\ref{fig_non extremal}), and the metric we used is not appropriate near the inner horizon.

On the other hand, one can confirm that there is no extremal surface at this early time limit when one attempts to find an extremal surface near the outer horizon. In this regard, the island region does not emerge at early time, and the entanglement entropy should be determined by (\ref{eq_entropy wo island}). In other words, the entanglement entropy grows linear in time at early time regime, which takes the following behavior

\begin{equation}
    S(R) = \frac{c}{3} \log{\left(2 \cosh{\kappa_+ t} \right)} \simeq \frac{c}{3} \kappa_+ t \,.
\end{equation}

\paragraph{Late time}
The late time behavior of the entanglement entropy is computed by taking the limit $r_+ < b \ll t_a,\, t_b$. In this limit, we use the following approximation

\begin{equation}
\cosh{\left(\kappa_+ t_{a,b}\right)} \simeq \frac{1}{2} \exp{\left(\kappa_+ t_{a,b}\right)}\,.
\end{equation}

Also, we employ the following approximation by assuming $a \simeq r_+$

\begin{equation}
    \log{\left[1-2\sqrt{\frac{a^2-r_+^2}{b^2}}\left(\frac{a^2-r_-^2}{b^2}\right)^{\frac{\kappa_+}{2\kappa_-}}\right]} \simeq -2\sqrt{\frac{a^2-r_+^2}{b^2}}\left(\frac{a^2-r_-^2}{b^2}\right)^{\frac{\kappa_+}{2\kappa_-}}.
\end{equation}

Using these approximations, one can write the generalized entropy as a sum of area term and the entanglement entropy of the matter fields
\begin{equation}\label{eq_gen entropy_non ext with island}
\begin{split}
    S_{\text{gen}} \simeq& \,\, \frac{a^2}{2 G_N} + \frac{c}{3} \left(2+\frac{\kappa_+}{\kappa_-}\right)\log{b} +\frac{c}{6}\log{\left[\frac{|a^2-r_-^2|^{1-\frac{\kappa_+}{\kappa_-}}}{\kappa_+^4 r_+^4 a^2}\right]} \\ &- \frac{2c}{3} \left(\frac{|a^2-r_+^2|}{b^2}\right)^\frac{1}{2} \left(\frac{|a^2-r_-^2|}{b^2}\right)^\frac{\kappa_+}{2 \kappa_-} \cosh{\left(\kappa_+(t_a-t_b)\right)} \,,
\end{split}
\end{equation}
where we take large $t_a$, $t_b$. The generalized entropy is extremized at $t_a = t_b$.

By extremizing (\ref{eq_gen entropy_non ext with island}) with respect to $a$, one can derive the location of the island
\begin{equation}\label{eq_location}
    a \simeq r_+ + \frac{2 c^2 G_N^2}{9 r_+ b^2}\left(\frac{r_+^2-r_-^2}{b^2}\right)^\frac{\kappa_+}{\kappa_-} \,.
\end{equation}
To find the extremal point, we assume that the island is located near the outer horizon $a \simeq r_+$ and expand the result by order of $G_N$. The result shows that the quantum extremal surface exists slightly outside the outer horizon at the late time regime. The existence of the island influences the behavior of the entanglement entropy for late time.

To see the effect of the island configuration, we insert the location of the island into the generalized entropy (\ref{eq_gen entropy_non ext with island}). The entanglement entropy is determined by
\begin{equation}\label{eq_w island final}
    S(R) \simeq \frac{r_+^2}{2 G_N} + \frac{c}{3}\left(2+\frac{\kappa_+}{\kappa_-}\right)\log{b}+\frac{c}{6} \log{\left[\frac{\left(r_+^2-r_-^2\right)^{-\frac{\kappa+}{\kappa_-}}}{\kappa_+^3 r_+^3}\right]}-\frac{2 c^2 G_N^2}{9 b^2}\left(\frac{r_+^2-r_-^2}{b^2}\right)^{\frac{\kappa+}{\kappa_-}} \,.
\end{equation}
Note that the entanglement entropy become constant at late times. At the leading order, the entanglement entropy reduces to the twice of Bekenstein-Hawking entropy, i.e.
\begin{equation}
    S(R) \approx 2 S_{BH} \,.
\end{equation}
 
The subleading terms contain the quantum correction of the entanglement entropy, which is not significant compared to the leading contribution of $S_{\text{BH}}$. This is what we can expect from the eternal black hole case explained in introduction.
When we take $Q \rightarrow 0$, our result is consistent with that in the neutral linear dilaton model~\cite{Karananas:2020fwx}. Based on their work\footnote{In order to check if \eqref{eq_w island final} can reproduce the neutral case result in \cite{Karananas:2020fwx}, one may need to compute the subleading terms of the entanglement entropy. However, the authors in \cite{Karananas:2020fwx} only presented the leading contribution. Thus, for the purpose of the comparison, we also compute $S(R)$ up to $c^2$-order using the presented formulae in \cite{Karananas:2020fwx}, which corresponds to \eqref{BFAV2}.}, we obtain the entanglement entropy for the neutral linear dilaton model up to second order of $c$, which can be reproduced from (\ref{eq_w island final}).

\begin{equation}\label{BFAV2}
    S(R) \simeq \frac{r_h^2}{2 G_N} + \frac{2c}{3} \log{b} - \frac{2 c^2 G_N}{9 b^2} \,,
\end{equation}
where $r_h$ is the horizon radius of the linear dilaton black hole.

\subsection{Page time and scrambling time}

\begin{figure}
\begin{center}
\setlength{\unitlength}{1cm}
\hspace{-0.9cm}
\includegraphics[width=10.5cm]{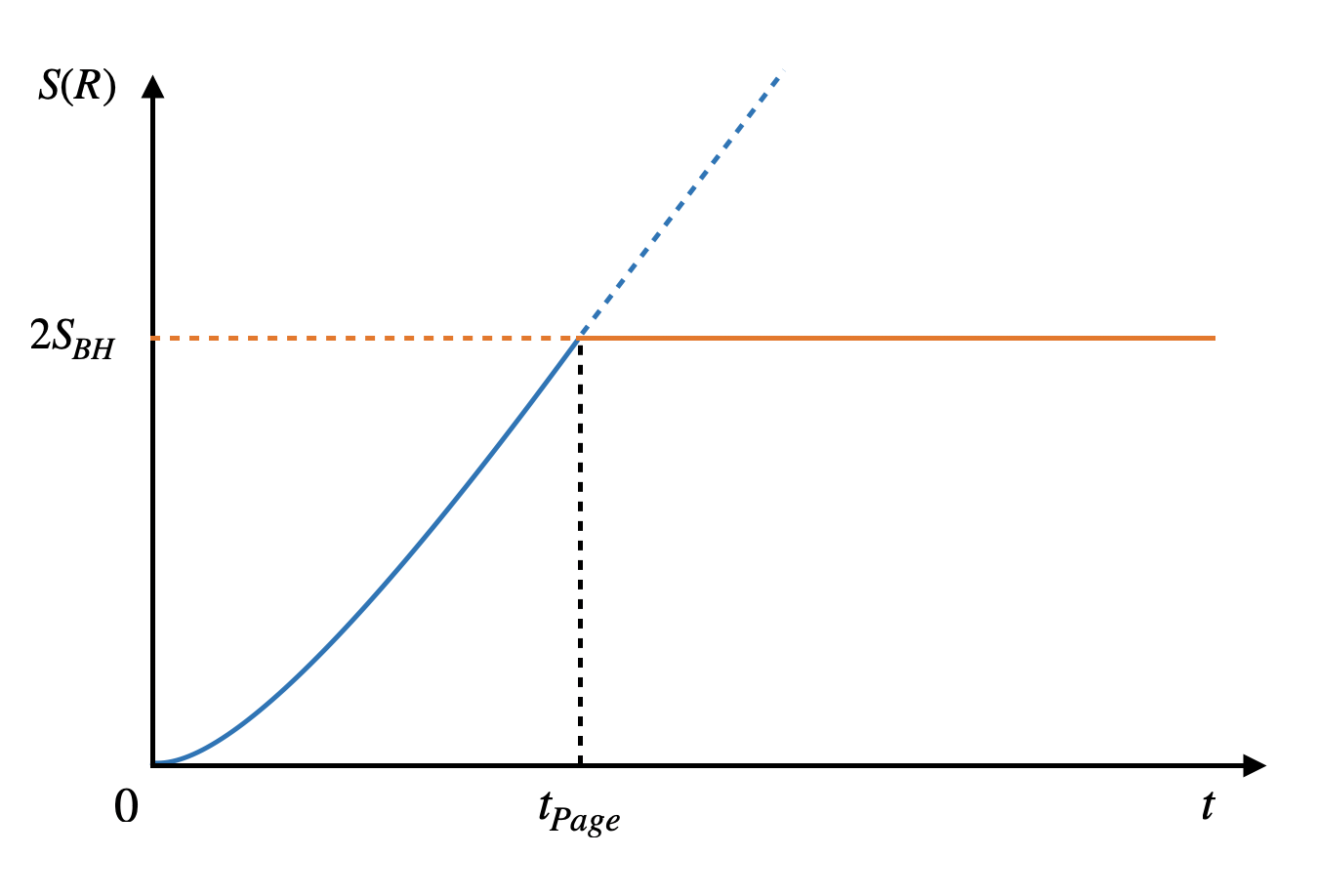}
\end{center}
\caption{The Page curve for the charged linear dilaton black holes. Without island configuration, the entanglement entropy grows in time (Blue dashed line). In the presence of the island, the entanglement entropy becomes constant at late times (Orange solid line).}
\label{fig_Page}
\end{figure}

In the previous section, we have observed that the entanglement entropy grows linearly in time at the early time regime. This linear growth is originated from the absence of the island. After converting into the late time regime, the island appears near the outer horizon. In this island phase, the entanglement entropy of the matter sector becomes constant. 

The transition between these two configurations can be described by the Page curve shown in Fig.~\ref{fig_Page}. In the Page curve, the Page time at which the linear growth become constant can be computed. By equating eqs. (\ref{eq_entropy wo island}) and (\ref{eq_w island final}), one can obtain
\begin{equation}
    t_{\text{Page}} = \frac{3 r_+^2}{2 \, c \, G_N \kappa_+} 
    = \frac{3}{ \pi c}\frac{S_{\text{BH}}}{T_H}\,, \label{paget}
\end{equation}
where we used the Hawking temperature $T_H=\kappa_+ / 2 \pi$. In this formula, one can confirm that the island arises around the Page time which is proportional to the Bekenstein-Hawking entropy of the black hole.

We can also consider the time scale for scrambling for the charged dilaton black hole. In the context of black hole information, the scrambling time is defined by the minimal time that one can retrieve the information after sending an information into the black hole. Note that the radiation degree of freedom is encoded in the union of two region $(R \cup \text{Island})$ with the presence of the island. In this case, the signal falling into the black hole come up in the radiation degree of freedom after the signal approach to the island. When the observer staying at the cutoff surface throws the signal into the black hole at time $t_0$, the signal will get to the island at the time $t_a$ that is given by
\begin{equation} \label{eq_t_a}
    t_{a}=t_{0}+\frac{1}{\kappa_+}\log{\left[\frac{|b-r_+|\,|b+r_+|}{|a-r_+|\,|a+r_+|}\right]}+\frac{1}{\kappa_-}\log{\left[\frac{|b-r_-|\,|b+r_-|}{|a-r_-|\,|a+r_-|}\right]}\,.
\end{equation}

Consider the case that the signal can be decoded from the Hawking radiation right after the signal reach the boundary of the island. In this situation, one can define the scrambling time by $t_{scr} = t_a - t_0$. In (\ref{eq_location}), the island is located near the event horizon $a \sim r_+ + \mathcal{O}\left((c \, G_N)^2/r_+^3\right)$. Then, the dominant term in eq. (\ref{eq_t_a}) yields the scrambling time as
\begin{equation}
    t_{\text{scr}} \simeq \frac{2}{\kappa_+}\log{\left(\frac{r_+^2}{G_N}\right)} \simeq \frac{1}{2 \pi T_H} \log{S_\text{BH}}\,,
\end{equation}
where we assumed that the central charge is much smaller than the Bekenstein-Hawking entropy, i.e. $c \ll S_{\text{BH}}$.

The scrambling time is comparable to logarithm of the Bekenstein-Hawking entropy. This is consistent with the argument of the fast scrambler in \cite{Sekino:2008he}. Our leading order computation says that the fast scrambling of the charged dilaton black holes can also be expected in the island prescription. Also, the decoding process associated with the scrambling time can be understood by the Hayden-Preskill protocol \cite{Hayden_2007}.

So far, we have considered non-extremal charged dilaton black holes. We have found the location of the boundary of island and by using it computed the entanglement entropy. It is interesting that these result can reproduce the Page curve that we expected for the unitary black holes. Also, the scrambling time can be derived from the island prescription.



\section{Entanglement entropy of extremal charged dilaton balck holes}

\begin{figure}
\begin{center}
    \subfigure[Without island]
    {\includegraphics[width=6cm]{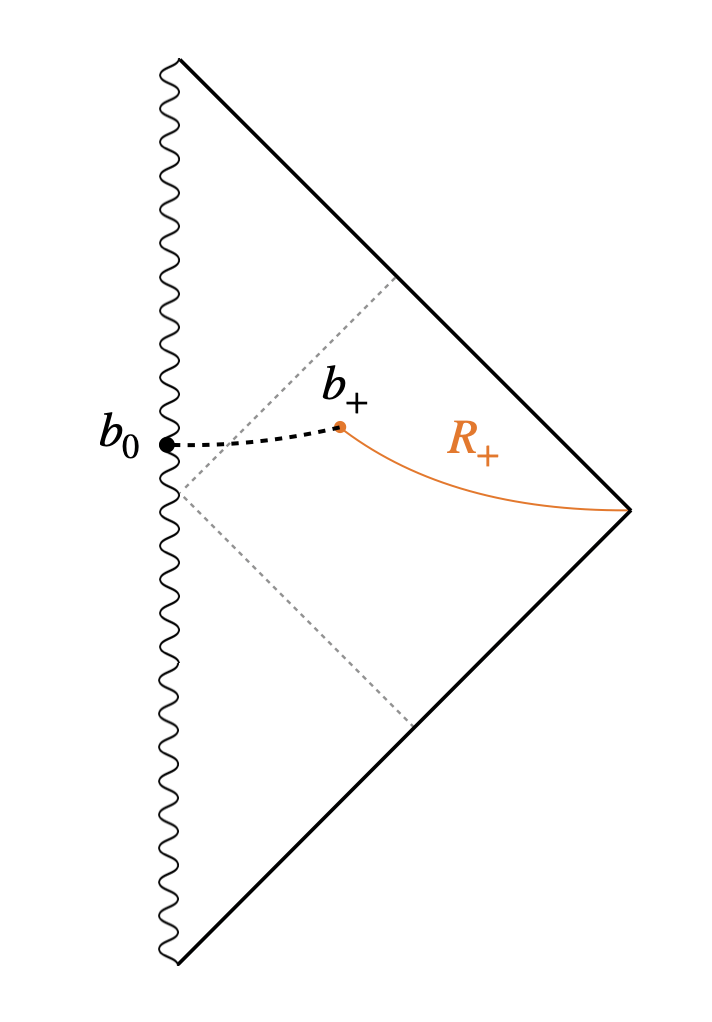}}
    \quad \quad
    \subfigure[With island]
    {\includegraphics[width=6cm]{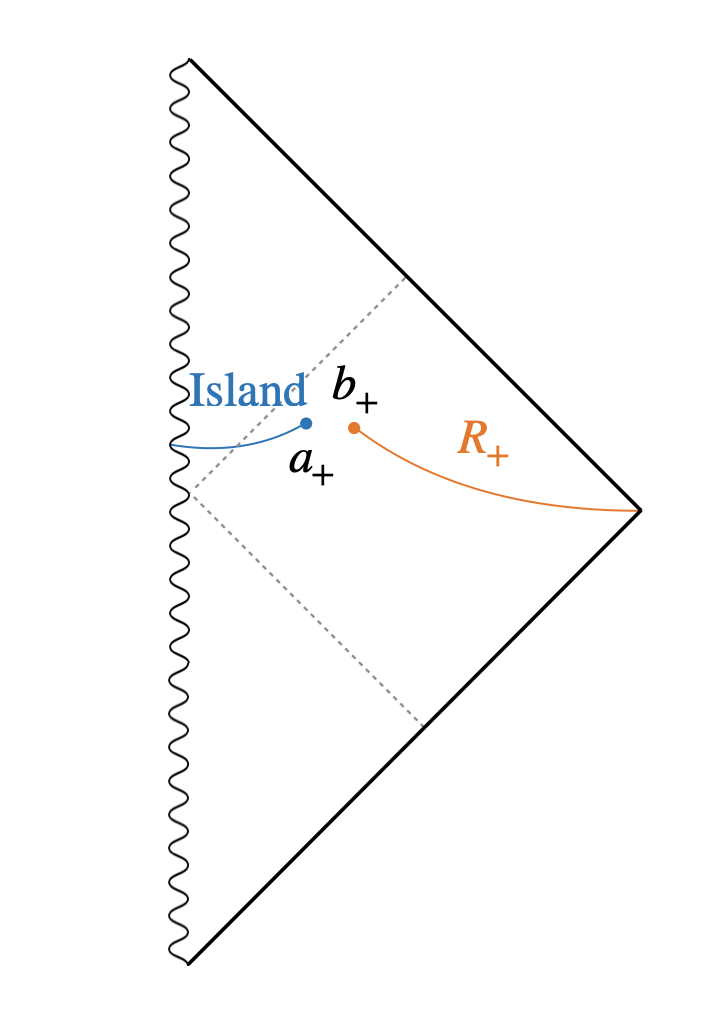}}
       
    \caption{(a) Penrose diagram of the extremal charged dilaton black holes without island. The radiation region is denoted by $R_+$. The cutoff surface is located at $b_+$. Note that the Cauchy surface hits the singularity $b_0$. (b) Penrose diagram of the extremal charged dilaton black holes with island. The island extends from $r=0$ to $a_+$.}
\label{fig_extremal}
\end{center}
\end{figure}

In this section, we revisit the computation on the entropy for the extremal charged linear dilaton black holes \cite{Karananas:2020fwx}. In \cite{Karananas:2020fwx}, 
the authors reported $S(R)$ without the island and the explicit computation of $S(R)$ in the presence of the island is not shown yet\footnote{Their main motivation is focused on the linear dilaton model without the charge.}. Also, $S(R)$ without the island is computed based on the Penrose diagram of the non-extremal case. However, because the Penrose diagram of the extremal case is not a continuous limit of the non-extremal case, we should start from the extremal setup.

One may suspect that considering $r_\pm = r_h \pm \epsilon$, the Penrose diagram of the non-extremal charged black holes shrinks to one of the extremal black holes, but it can not happen. It is not a matter of how close two horizons are, but of what a causal structure is.
Even though we start from the Penrose diagram of the extremal black holes, the Cauchy surface can not avoid meeting the singularity~\cite{kim2021entanglement}. By carefully considering the Penrose diagram, we calculate the entanglement entropy of the extremal charged linear dilaton black holes.



\kyr{
}

\subsection{Entropy without island}

As in Fig.~\ref{fig_extremal}, the Cauchy surface including $b_+ = (t_b,b)$ touches the singularity at $b_0 = (t_b,0)$. In this case, the entropy of the matter field is given by the geodesic distance between $b_+$ and $b_0$. By using (\ref{eq_wo island}), the entanglement entropy of matter field is given by
\begin{align}
    S_{\text{matter}}&=\frac{c}{3}\log{l(b_+,b_0)}\\
    &\sim \frac{c}{6} \log{\left[f(b_+)f(b_0)(u(b_0)-u(b_+))(v(b_+)-v(b_0))\right]}\,. \nonumber
\end{align}\label{EWOIEXT}

However, since the line element in the Kruskal coordinate has singular point at $r=r_h$, the expression of the geodesic distance between $b_+$ and $b_0$ is ambiguous in the second line. Also, one can easily check that the $f(b_0)$ is ill-defined: from \eqref{ILLNEESNEEDSTOBECURE}, one can find that $f(0)$ can not be evaluated due to the divergence at $r=0$. This fact makes it difficult to compute the entropy of the matter sector for the extremal case.

\subsection{Entropy with island}
In the presence of the island, one need to compute the geodesic distance between $a_+ = (t_a, a)$ and $b_+ = (t_b, b)$ in Fig.~\ref{fig_extremal} to obtain the entropy of radiation. This is because that one can avoid a difficulty of singularity and the geodesic distance $l(a_+,b_+)$ is well-defined. Using the Kruskal coordinates for the extemal case in section \ref{section2}, one can obtain
\begin{equation}
    S_{\text{matter}} = \frac{c}{3} \log{\left[l(a_+,b_+)\right]} = \frac{c}{6} \log A + \frac{c}{6} \log \left[B + \frac{1}{B} - 2 \cosh{\left(\frac{t_a-t_b}{r_h}\right)}\right]\,,
\end{equation}
where
\begin{equation}
    \begin{split}
        A &= \frac{(a^2-r_h^2) (b^2-r_h^2)}{a b}\,,\\
        B &= \sqrt{\frac{b^2-r_h^2}{a^2-r_h^2}} \exp{\left[\frac{1}{2}\left(\frac{r_h^2}{a^2-r_h^2}-\frac{r_h^2}{b^2-r_h^2}\right)\right]}\,.
    \end{split}
\end{equation}
To compute the entropy of the matter field, we assume that the cutoff surface locates far from the horizon $b \gg r_h$
\begin{equation}
\begin{split}
    A &\simeq \frac{b}{a}(a^2-r_h^2)\,, \\
    B &\simeq \sqrt{\frac{b^2}{a^2-r_h^2}}\exp{\left[\frac{1}{2}\left(\frac{r_h^2}{a^2-r_h^2}\right)\right]}\,.
\end{split}
\end{equation}

By adding the area term originated from the island contribution, the generalized entropy is given by
\begin{equation} \label{eq_ext_gen}
\begin{split}
    S_{\text{gen}} \simeq& \frac{a^2}{4 G_N} +\frac{c}{6}\log{\left[\sqrt{\frac{b^2}{a^2-r_h^2}}\exp{\left[\frac{1}{2}\left(\frac{r_h^2}{a^2-r_h^2}\right)\right]}+\sqrt{\frac{a^2-r_h^2}{b^2}}\exp{\left[-\frac{1}{2}\left(\frac{r_h^2}{a^2-r_h^2}\right)\right]}\right]} \\ 
    &+ \frac{c}{6} \log{\left[\frac{b}{a}(a^2-r_h^2)\right]}-\frac{2c}{3}\sqrt{\frac{a^2-r_h^2}{b^2}}\exp{\left[-\frac{1}{2}\left(\frac{r_h^2}{a^2-r_h^2}\right)\right]}\sinh^2{\left(\frac{t_a-t_b}{2r_h}\right)}\,.
\end{split}
\end{equation}

The generalized entropy (\ref{eq_ext_gen}) has an extremal value at $t_a = t_b$. To extremize the generalized entropy with respect to $a$, we consider the fact that the cutoff surface is located far from the event horizon. Then, one can find the location of the extremal surface:
\begin{equation}
    a \simeq r_h +\sqrt{\frac{c}{12}} \, l_P\,,
\end{equation}
where $l_P$ is the Planck length, and we used $a \simeq r_h$.
By inserting $a$ into the generalized entropy, one can derive the entanglement entropy of radiation in the extremal case
\begin{equation}
    S(R) \simeq \frac{r_h^2}{4 G_N} +\sqrt{\frac{c}{12}} \frac{r_h}{l_P}+ \mathcal{O}(c)\,.
\end{equation}

The entanglement entropy is constant with island configuration even in the extremal case. It is noteworthy that the entanglement entropy of the extremal charged dilaton case is comparable to the Bekenstein-Hawking entropy: 
\begin{equation}
    S(R) \approx  S_{BH} \,.
\end{equation}
This result seems natural becuase the causal structure of the extremal charged case forms one-side black hole (see Fig.~\ref{fig_extremal}).

Note that the entanglement entropy for the extremal charged dilaton black hole can not be obtained by the countinuous limit from the non-extremal charged dilaton balck hole. This caveat is originated from the difference of causal structure, which causes a different appearance of island in our model. In addition, due to the different causal structure, we could not find the linear growth of the entanglement entropy in the early time phase. As a consequence, we can not derive the Page time in the extremal case.
The difficulty of the continuous extremal limit in the entropy has also been discussed in \cite{kim2021entanglement, Carroll:2009maa}.\footnote{In \cite{Yu:2021cgi} the authors show that the charged dilatonic black hole gives the divergent or vanishing Page time for the extremal case. However, it seems that their extremal case is considered as the continuous limit of the non-extremal case, which is different from our method.
}


In summary, the charged linear dilaton model is the natural extension and the complementary to the previous works in many aspects explained above.

%
\section{Conclusions}
We have studied the entanglement entropy of the Hawking radiation, $S(R)$, of the eternal black hole using the island formula \eqref{ISFOR}. In particular, our work aims to investigate the information paradox with the Page curve of the 4 dimensional charged linear dilaton model \eqref{ACTION} with two main motivations.
\paragraph{Motivation 1: why the linear dilaton model?}
it would be important to check the range of applicability of the island method if it can be applied to all kinds of black hole geometries. For this purpose, the linear dilaton model could be one of possible candidates in that it allows a metric not asymptotically flat/AdS/dS, which is a non-standard black hole geometry. Note that this first motivation is significant not only for the applicability perspective, but also for the better understanding of the quantum gravity for the general black holes.
\paragraph{Motivation 2: why we need to consider the charge on it?}
the action \eqref{ACTION} can contribute to obtain more complete picture of the Page curve in that it has both the non-extremal black hole and the extremal black hole because of the finite charge. Most of the literatures considered only for the non-extremal black holes and there has been reported, such as the asymptotically flat black hole case~\cite{kim2021entanglement,Yu:2021cgi}, that the extremal case produces the different result from non-extremal case.
Thus we have studied the action \eqref{ACTION} because it may help to better understand the Page curve of the non-standard black holes in more complete framework including extremal black holes. 

\paragraph{The non-extremal black hole:}
From the separate analysis of the non-extremal black holes and the extremal black hole, we found that the island formula works for the non-extremal black holes, i.e., $S(R)$ grows linearly in time without the island and it is bounded by double of the Bekenstein-Hawking entropy ($2 S_{\text{BH}}$) in the presence of the island, which is consistent with the Page curve incorporating with what the unitarity principle requires. Moreover we also derived the Page time and the scrambling time consistent with the Hayden-Preskill protocol.

\paragraph{The extremal black hole:}
For the extremal black holes, it turns out that $S(R)$ without the island produces the ill-defined result. 
However, $S(R)$ in the presence of the island can provide the finite entropy as $S_{\text{BH}}$. Note that the result of the extremal case ($S(R) \sim S_{\text{BH}}$) can not be obtained from the continuous extremal limit of the non-extremal black hole case ($S(R) \sim 2S_{\text{BH}}$). Note also that the origin of the differences between the non-extremal black holes and the extremal black holes comes from the fact that the Penrose diagram of them are different.

\begin{figure}[]
\centering
     {\includegraphics[width=15.2cm]{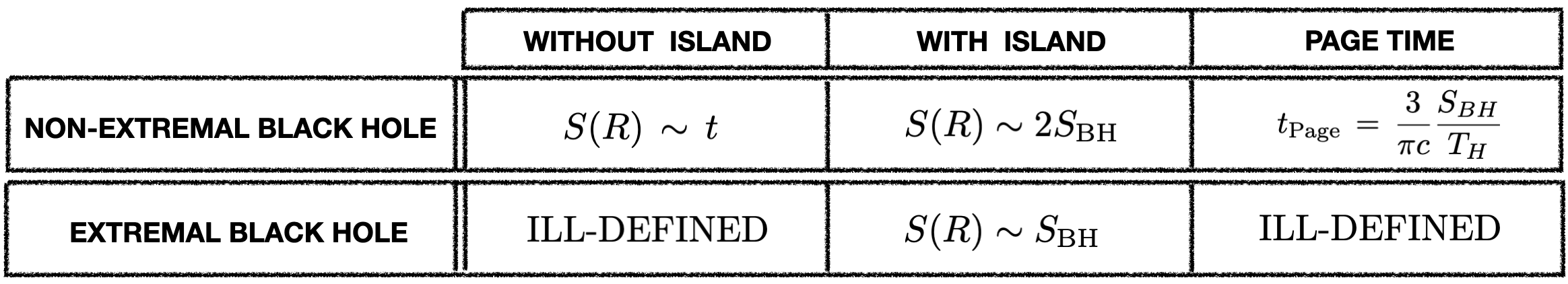} \label{}}
 \caption{The summary of results. Unlike the non-extremal black holes, the island formula can not produce the Page curve for the extremal black holes.}\label{TBFd1}
\end{figure}

In Fig. \ref{TBFd1}, we make the summary table of our results.
We further make a few more comments on what we found. 
First, using the same action \eqref{ACTION}, there was a previous study of $S(R)$ for the extremal case~\cite{Karananas:2020fwx} and they argued: i) $S(R)$ without the island can provide a well defined quantity; ii) $S(R)$ with the island behaves as $2S_{\text{BH}}$ even for the extremal black hole. These are different from what we found. This disagreement appears because in \cite{Karananas:2020fwx} the generalized entropy of the extremal case is evaluated  with the formula for the non-extremal black hole. In our paper, because the Penrose diagram of the extremal case is not a continuous limit of the non-extremal case, we start from the extremal setup directly.



Second, our results can be compared with the asymptotically flat black holes studies for the extremal black holes~\cite{kim2021entanglement,Yu:2021cgi}.
For the Reissner-Nordstrom black 
holes~\cite{kim2021entanglement}, the authors also reported the same results presented in Fig. \ref{TBFd1}.
On the other hand, for the Garfinkle-Horowitz-Strominger dilaton black holes (the charged dilaton black holes)~\cite{Yu:2021cgi}, the authors argued the Page time is vanishing or divergent for the extremal case, which is different from our result and \cite{kim2021entanglement}.
This disagreement again is related to the way to study the extremal black holes: 
their result is concluded from the extremal limit of the non-extremal black hole results. One may expect that once the extremal black hole analysis is separately performed with the Penrose diagram like Fig. \ref{fig_extremal}, the Garfinkle-Horowitz-Strominger dilaton black holes may also produce the similar results in Fig. \ref{TBFd1}.
Inspired from this work, it would be interesting to investigate if the table in Fig. \ref{TBFd1} could be universal feature for all the black hole geometries at finite charges.

Third, note that the page time \eqref{paget} is universal for all different models~\cite{Hashimoto:2020cas,Karananas:2020fwx,Wang:2021woy,kim2021entanglement,Yu:2021cgi} studied by our method:
\begin{equation*}
    t_{\text{Page}}
    = \frac{3}{ \pi c}\frac{S_{\text{BH}}}{T_H}\,. 
\end{equation*}
Note also that one can not tell the Page times for extremal cases are vanishing (or divergent) unless $S(R)$ is well defined.

Another interesting future direction would be considering the case with more than one island. In our work, we have studied only the case with zero island or one island. However, in general, it would be possible to see the effect of the configuration with several islands on the entanglement entropy, for instance the multiple islands may soften the sharp change of the Page curve at the Page time.
In addition to the configuration of the islands, there is the open question about the island from the information perspective: how the information in the island is transformed into the radiation region? Although the Page curve in the unitary fashion is reproduced, it can not tell the mechanism behind how the information leaks out from the island and transformed. One intriguing argument to explain this leakage of the information is ER$=$EPR~\cite{Maldacena:2013xja}, although the more formal mathematical proof is needed. 
We hope to address these questions in the near future.

\acknowledgments

We would like to thank  Mitsuhiro Nishida for valuable discussions and correspondence.  
This work was supported by the National Key R$\&$D Program of China (Grant No. 2018FYA0305800), Project 12035016 supported by National Natural Science Foundation of China, the Strategic Priority Research Program of Chinese Academy of Sciences, Grant No. XDB28000000, Basic Science Research Program through the National Research Foundation of Korea (NRF) funded by the Ministry of Science, ICT $\&$ Future Planning (NRF- 2021R1A2C1006791) and GIST Research Institute(GRI) grant funded by the GIST in 2021. B. Ahn was also supported by Basic Science Research Program through the National Research Foundation of Korea funded by the Ministry of Education (NRF-2020R1A6A3A01095962).

\appendix
%



\begin{thebibliography}{10}

\bibitem{Hawking:1976aa}
S.~W. Hawking, \emph{Breakdown of predictability in gravitational collapse},
  \href{http://dx.doi.org/10.1103/PhysRevD.14.2460}{\emph{Physical Review D}
  {\bf 14} (1976) 2460--2473}.

\bibitem{Hawking_1975}
S.~W. Hawking, \emph{Particle creation by black holes},
  \href{http://dx.doi.org/10.1007/bf02345020}{\emph{Communications In
  Mathematical Physics} {\bf 43} (aug, 1975) 199--220}.

\bibitem{Page:1993wv}
D.~N. Page, \emph{{Information in black hole radiation}},
  \href{http://dx.doi.org/10.1103/PhysRevLett.71.3743}{\emph{Phys. Rev. Lett.}
  {\bf 71} (1993) 3743--3746}, [\href{http://arxiv.org/abs/hep-th/9306083}{{\tt
  hep-th/9306083}}].

\bibitem{Wald_1975}
R.~M. Wald, \emph{On particle creation by black holes},
  \href{http://dx.doi.org/10.1007/bf01609863}{\emph{Communications in
  Mathematical Physics} {\bf 45} (feb, 1975) 9--34}.

\bibitem{Parker:1975aa}
L.~Parker, \emph{Probability distribution of particles created by a black
  hole}, \href{http://dx.doi.org/10.1103/PhysRevD.12.1519}{\emph{Physical
  Review D} {\bf 12} (1975) 1519--1525}.

\bibitem{Almheiri:2019hni}
A.~Almheiri, R.~Mahajan, J.~Maldacena and Y.~Zhao, \emph{{The Page curve of
  Hawking radiation from semiclassical geometry}},
  \href{http://dx.doi.org/10.1007/JHEP03(2020)149}{\emph{JHEP} {\bf 03} (2020)
  149}, [\href{http://arxiv.org/abs/1908.10996}{{\tt 1908.10996}}].

\bibitem{Almheiri:2019yqk}
A.~Almheiri, R.~Mahajan and J.~Maldacena, \emph{{Islands outside the horizon}},
   \href{http://arxiv.org/abs/1910.11077}{{\tt 1910.11077}}.

\bibitem{Almheiri:2019qdq}
A.~Almheiri, T.~Hartman, J.~Maldacena, E.~Shaghoulian and A.~Tajdini,
  \emph{{Replica Wormholes and the Entropy of Hawking Radiation}},
  \href{http://dx.doi.org/10.1007/JHEP05(2020)013}{\emph{JHEP} {\bf 05} (2020)
  013}, [\href{http://arxiv.org/abs/1911.12333}{{\tt 1911.12333}}].

\bibitem{Almheiri:2019psf}
A.~Almheiri, N.~Engelhardt, D.~Marolf and H.~Maxfield, \emph{{The entropy of
  bulk quantum fields and the entanglement wedge of an evaporating black
  hole}}, \href{http://dx.doi.org/10.1007/JHEP12(2019)063}{\emph{JHEP} {\bf 12}
  (2019) 063}, [\href{http://arxiv.org/abs/1905.08762}{{\tt 1905.08762}}].

\bibitem{Penington:2019npb}
G.~Penington, \emph{{Entanglement Wedge Reconstruction and the Information
  Paradox}}, \href{http://dx.doi.org/10.1007/JHEP09(2020)002}{\emph{JHEP} {\bf
  09} (2020) 002}, [\href{http://arxiv.org/abs/1905.08255}{{\tt 1905.08255}}].

\bibitem{Engelhardt:2014gca}
N.~Engelhardt and A.~C. Wall, \emph{{Quantum Extremal Surfaces: Holographic
  Entanglement Entropy beyond the Classical Regime}},
  \href{http://dx.doi.org/10.1007/JHEP01(2015)073}{\emph{JHEP} {\bf 01} (2015)
  073}, [\href{http://arxiv.org/abs/1408.3203}{{\tt 1408.3203}}].

\bibitem{Bekenstein:1981aa}
J.~D. Bekenstein, \emph{Universal upper bound on the entropy-to-energy ratio
  for bounded systems},
  \href{http://dx.doi.org/10.1103/PhysRevD.23.287}{\emph{Physical Review D}
  {\bf 23} (1981) 287--298}.

\bibitem{Page:2013dx}
D.~N. Page, \emph{{Time Dependence of Hawking Radiation Entropy}},
  \href{http://dx.doi.org/10.1088/1475-7516/2013/09/028}{\emph{JCAP} {\bf 09}
  (2013) 028}, [\href{http://arxiv.org/abs/1301.4995}{{\tt 1301.4995}}].

\bibitem{Page:1993df}
D.~N. Page, \emph{{Average entropy of a subsystem}},
  \href{http://dx.doi.org/10.1103/PhysRevLett.71.1291}{\emph{Phys. Rev. Lett.}
  {\bf 71} (1993) 1291--1294}, [\href{http://arxiv.org/abs/gr-qc/9305007}{{\tt
  gr-qc/9305007}}].

\bibitem{Akers:2019lzs}
C.~Akers, N.~Engelhardt, G.~Penington and M.~Usatyuk, \emph{{Quantum Maximin
  Surfaces}}, \href{http://dx.doi.org/10.1007/JHEP08(2020)140}{\emph{JHEP} {\bf
  08} (2020) 140}, [\href{http://arxiv.org/abs/1912.02799}{{\tt 1912.02799}}].

\bibitem{Faulkner:2013ana}
T.~Faulkner, A.~Lewkowycz and J.~Maldacena, \emph{{Quantum corrections to
  holographic entanglement entropy}},
  \href{http://dx.doi.org/10.1007/JHEP11(2013)074}{\emph{JHEP} {\bf 11} (2013)
  074}, [\href{http://arxiv.org/abs/1307.2892}{{\tt 1307.2892}}].

\bibitem{Wall:2012uf}
A.~C. Wall, \emph{{Maximin Surfaces, and the Strong Subadditivity of the
  Covariant Holographic Entanglement Entropy}},
  \href{http://dx.doi.org/10.1088/0264-9381/31/22/225007}{\emph{Class. Quant.
  Grav.} {\bf 31} (2014) 225007}, [\href{http://arxiv.org/abs/1211.3494}{{\tt
  1211.3494}}].

\bibitem{Page:1979tc}
D.~N. Page, \emph{{IS BLACK HOLE EVAPORATION PREDICTABLE?}},
  \href{http://dx.doi.org/10.1103/PhysRevLett.44.301}{\emph{Phys. Rev. Lett.}
  {\bf 44} (1980) 301}.

\bibitem{Ryu:2006bv}
S.~Ryu and T.~Takayanagi, \emph{{Holographic derivation of entanglement entropy
  from AdS/CFT}},
  \href{http://dx.doi.org/10.1103/PhysRevLett.96.181602}{\emph{Phys. Rev.
  Lett.} {\bf 96} (2006) 181602},
  [\href{http://arxiv.org/abs/hep-th/0603001}{{\tt hep-th/0603001}}].

\bibitem{Hubeny:2007xt}
V.~E. Hubeny, M.~Rangamani and T.~Takayanagi, \emph{{A Covariant holographic
  entanglement entropy proposal}},
  \href{http://dx.doi.org/10.1088/1126-6708/2007/07/062}{\emph{JHEP} {\bf 07}
  (2007) 062}, [\href{http://arxiv.org/abs/0705.0016}{{\tt 0705.0016}}].

\bibitem{Lewkowycz:2013nqa}
A.~Lewkowycz and J.~Maldacena, \emph{{Generalized gravitational entropy}},
  \href{http://dx.doi.org/10.1007/JHEP08(2013)090}{\emph{JHEP} {\bf 08} (2013)
  090}, [\href{http://arxiv.org/abs/1304.4926}{{\tt 1304.4926}}].

\bibitem{Barrella:2013wja}
T.~Barrella, X.~Dong, S.~A. Hartnoll and V.~L. Martin, \emph{{Holographic
  entanglement beyond classical gravity}},
  \href{http://dx.doi.org/10.1007/JHEP09(2013)109}{\emph{JHEP} {\bf 09} (2013)
  109}, [\href{http://arxiv.org/abs/1306.4682}{{\tt 1306.4682}}].

\bibitem{Penington:2019kki}
G.~Penington, S.~H. Shenker, D.~Stanford and Z.~Yang, \emph{{Replica wormholes
  and the black hole interior}},  \href{http://arxiv.org/abs/1911.11977}{{\tt
  1911.11977}}.

\bibitem{Hartman:2020swn}
T.~Hartman, E.~Shaghoulian and A.~Strominger, \emph{{Islands in Asymptotically
  Flat 2D Gravity}},
  \href{http://dx.doi.org/10.1007/JHEP07(2020)022}{\emph{JHEP} {\bf 07} (2020)
  022}, [\href{http://arxiv.org/abs/2004.13857}{{\tt 2004.13857}}].

\bibitem{Goto:2020wnk}
K.~Goto, T.~Hartman and A.~Tajdini, \emph{{Replica wormholes for an evaporating
  2D black hole}}, \href{http://dx.doi.org/10.1007/JHEP04(2021)289}{\emph{JHEP}
  {\bf 04} (2021) 289}, [\href{http://arxiv.org/abs/2011.09043}{{\tt
  2011.09043}}].

\bibitem{Almheiri:2019psy}
A.~Almheiri, R.~Mahajan and J.~E. Santos, \emph{{Entanglement islands in higher
  dimensions}},
  \href{http://dx.doi.org/10.21468/SciPostPhys.9.1.001}{\emph{SciPost Phys.}
  {\bf 9} (2020) 001}, [\href{http://arxiv.org/abs/1911.09666}{{\tt
  1911.09666}}].

\bibitem{Hashimoto:2020cas}
K.~Hashimoto, N.~Iizuka and Y.~Matsuo, \emph{{Islands in Schwarzschild black
  holes}}, \href{http://dx.doi.org/10.1007/JHEP06(2020)085}{\emph{JHEP} {\bf
  06} (2020) 085}, [\href{http://arxiv.org/abs/2004.05863}{{\tt 2004.05863}}].

\bibitem{Karananas:2020fwx}
G.~K. Karananas, A.~Kehagias and J.~Taskas, \emph{{Islands in linear dilaton
  black holes}}, \href{http://dx.doi.org/10.1007/JHEP03(2021)253}{\emph{JHEP}
  {\bf 03} (2021) 253}, [\href{http://arxiv.org/abs/2101.00024}{{\tt
  2101.00024}}].

\bibitem{Wang:2021woy}
X.~Wang, R.~Li and J.~Wang, \emph{{Islands and Page curves of
  Reissner-Nordstr\"om black holes}},
  \href{http://dx.doi.org/10.1007/JHEP04(2021)103}{\emph{JHEP} {\bf 04} (2021)
  103}, [\href{http://arxiv.org/abs/2101.06867}{{\tt 2101.06867}}].

\bibitem{Yu:2021cgi}
M.-H. Yu and X.-H. Ge, \emph{{Page Curves and Islands in Charged Dilaton Black
  Holes}},  \href{http://arxiv.org/abs/2107.03031}{{\tt 2107.03031}}.

\bibitem{Alishahiha:2020qza}
M.~Alishahiha, A.~Faraji~Astaneh and A.~Naseh, \emph{{Island in the presence of
  higher derivative terms}},
  \href{http://dx.doi.org/10.1007/JHEP02(2021)035}{\emph{JHEP} {\bf 02} (2021)
  035}, [\href{http://arxiv.org/abs/2005.08715}{{\tt 2005.08715}}].

\bibitem{Geng:2020qvw}
H.~Geng and A.~Karch, \emph{{Massive islands}},
  \href{http://dx.doi.org/10.1007/JHEP09(2020)121}{\emph{JHEP} {\bf 09} (2020)
  121}, [\href{http://arxiv.org/abs/2006.02438}{{\tt 2006.02438}}].

\bibitem{almheiri2019islands}
A.~Almheiri, R.~Mahajan and J.~Maldacena, \emph{Islands outside the horizon},
  \href{http://arxiv.org/abs/1910.11077}{{\tt 1910.11077}}.

\bibitem{kim2021entanglement}
W.~Kim and M.~Nam, \emph{Entanglement entropy of asymptotically flat
  non-extremal and extremal black holes with an island},
  \href{http://arxiv.org/abs/2103.16163}{{\tt 2103.16163}}.

\bibitem{Almheiri:2020cfm}
A.~Almheiri, T.~Hartman, J.~Maldacena, E.~Shaghoulian and A.~Tajdini,
  \emph{{The entropy of Hawking radiation}},
  \href{http://arxiv.org/abs/2006.06872}{{\tt 2006.06872}}.

\bibitem{Chen:2019uhq}
H.~Z. Chen, Z.~Fisher, J.~Hernandez, R.~C. Myers and S.-M. Ruan,
  \emph{{Information Flow in Black Hole Evaporation}},
  \href{http://dx.doi.org/10.1007/JHEP03(2020)152}{\emph{JHEP} {\bf 03} (2020)
  152}, [\href{http://arxiv.org/abs/1911.03402}{{\tt 1911.03402}}].

\bibitem{Chen:2019iro}
Y.~Chen, \emph{{Pulling Out the Island with Modular Flow}},
  \href{http://dx.doi.org/10.1007/JHEP03(2020)033}{\emph{JHEP} {\bf 03} (2020)
  033}, [\href{http://arxiv.org/abs/1912.02210}{{\tt 1912.02210}}].

\bibitem{Gautason:2020tmk}
F.~F. Gautason, L.~Schneiderbauer, W.~Sybesma and L.~Thorlacius, \emph{{Page
  Curve for an Evaporating Black Hole}},
  \href{http://dx.doi.org/10.1007/JHEP05(2020)091}{\emph{JHEP} {\bf 05} (2020)
  091}, [\href{http://arxiv.org/abs/2004.00598}{{\tt 2004.00598}}].

\bibitem{Anegawa:2020ezn}
T.~Anegawa and N.~Iizuka, \emph{{Notes on islands in asymptotically flat 2d
  dilaton black holes}},
  \href{http://dx.doi.org/10.1007/JHEP07(2020)036}{\emph{JHEP} {\bf 07} (2020)
  036}, [\href{http://arxiv.org/abs/2004.01601}{{\tt 2004.01601}}].

\bibitem{Hollowood:2020cou}
T.~J. Hollowood and S.~P. Kumar, \emph{{Islands and Page Curves for Evaporating
  Black Holes in JT Gravity}},
  \href{http://dx.doi.org/10.1007/JHEP08(2020)094}{\emph{JHEP} {\bf 08} (2020)
  094}, [\href{http://arxiv.org/abs/2004.14944}{{\tt 2004.14944}}].

\bibitem{Krishnan:2020oun}
C.~Krishnan, V.~Patil and J.~Pereira, \emph{{Page Curve and the Information
  Paradox in Flat Space}},  \href{http://arxiv.org/abs/2005.02993}{{\tt
  2005.02993}}.

\bibitem{Banks:2020zrt}
T.~Banks, \emph{{Microscopic Models of Linear Dilaton Gravity and Their
  Semi-classical Approximations}},  \href{http://arxiv.org/abs/2005.09479}{{\tt
  2005.09479}}.

\bibitem{Chen:2020uac}
H.~Z. Chen, R.~C. Myers, D.~Neuenfeld, I.~A. Reyes and J.~Sandor,
  \emph{{Quantum Extremal Islands Made Easy, Part I: Entanglement on the
  Brane}}, \href{http://dx.doi.org/10.1007/JHEP10(2020)166}{\emph{JHEP} {\bf
  10} (2020) 166}, [\href{http://arxiv.org/abs/2006.04851}{{\tt 2006.04851}}].

\bibitem{Chandrasekaran:2020qtn}
V.~Chandrasekaran, M.~Miyaji and P.~Rath, \emph{{Including contributions from
  entanglement islands to the reflected entropy}},
  \href{http://dx.doi.org/10.1103/PhysRevD.102.086009}{\emph{Phys. Rev. D} {\bf
  102} (2020) 086009}, [\href{http://arxiv.org/abs/2006.10754}{{\tt
  2006.10754}}].

\bibitem{Li:2020ceg}
T.~Li, J.~Chu and Y.~Zhou, \emph{{Reflected Entropy for an Evaporating Black
  Hole}}, \href{http://dx.doi.org/10.1007/JHEP11(2020)155}{\emph{JHEP} {\bf 11}
  (2020) 155}, [\href{http://arxiv.org/abs/2006.10846}{{\tt 2006.10846}}].

\bibitem{Bak:2020enw}
D.~Bak, C.~Kim, S.-H. Yi and J.~Yoon, \emph{{Unitarity of entanglement and
  islands in two-sided Janus black holes}},
  \href{http://dx.doi.org/10.1007/JHEP01(2021)155}{\emph{JHEP} {\bf 01} (2021)
  155}, [\href{http://arxiv.org/abs/2006.11717}{{\tt 2006.11717}}].

\bibitem{Bousso:2020kmy}
R.~Bousso and E.~Wildenhain, \emph{{Gravity/ensemble duality}},
  \href{http://dx.doi.org/10.1103/PhysRevD.102.066005}{\emph{Phys. Rev. D} {\bf
  102} (2020) 066005}, [\href{http://arxiv.org/abs/2006.16289}{{\tt
  2006.16289}}].

\bibitem{Hollowood:2020kvk}
T.~J. Hollowood, S.~Prem~Kumar and A.~Legramandi, \emph{{Hawking radiation
  correlations of evaporating black holes in JT gravity}},
  \href{http://dx.doi.org/10.1088/1751-8121/abbc51}{\emph{J. Phys. A} {\bf 53}
  (2020) 475401}, [\href{http://arxiv.org/abs/2007.04877}{{\tt 2007.04877}}].

\bibitem{Krishnan:2020fer}
C.~Krishnan, \emph{{Critical Islands}},
  \href{http://dx.doi.org/10.1007/JHEP01(2021)179}{\emph{JHEP} {\bf 01} (2021)
  179}, [\href{http://arxiv.org/abs/2007.06551}{{\tt 2007.06551}}].

\bibitem{Engelhardt:2020qpv}
N.~Engelhardt, S.~Fischetti and A.~Maloney, \emph{{Free energy from replica
  wormholes}}, \href{http://dx.doi.org/10.1103/PhysRevD.103.046021}{\emph{Phys.
  Rev. D} {\bf 103} (2021) 046021},
  [\href{http://arxiv.org/abs/2007.07444}{{\tt 2007.07444}}].

\bibitem{Karlsson:2020uga}
A.~Karlsson, \emph{{Replica wormhole and island incompatibility with monogamy
  of entanglement}},  \href{http://arxiv.org/abs/2007.10523}{{\tt 2007.10523}}.

\bibitem{Gomez:2020yef}
C.~Gomez, \emph{{The information of the information paradox: On the quantum
  information meaning of Page curve}},
  \href{http://arxiv.org/abs/2007.11508}{{\tt 2007.11508}}.

\bibitem{Chen:2020jvn}
H.~Z. Chen, Z.~Fisher, J.~Hernandez, R.~C. Myers and S.-M. Ruan,
  \emph{{Evaporating Black Holes Coupled to a Thermal Bath}},
  \href{http://dx.doi.org/10.1007/JHEP01(2021)065}{\emph{JHEP} {\bf 01} (2021)
  065}, [\href{http://arxiv.org/abs/2007.11658}{{\tt 2007.11658}}].

\bibitem{Hartman:2020khs}
T.~Hartman, Y.~Jiang and E.~Shaghoulian, \emph{{Islands in cosmology}},
  \href{http://dx.doi.org/10.1007/JHEP11(2020)111}{\emph{JHEP} {\bf 11} (2020)
  111}, [\href{http://arxiv.org/abs/2008.01022}{{\tt 2008.01022}}].

\bibitem{Balasubramanian:2020coy}
V.~Balasubramanian, A.~Kar and T.~Ugajin, \emph{{Entanglement between two
  disjoint universes}},
  \href{http://dx.doi.org/10.1007/JHEP02(2021)136}{\emph{JHEP} {\bf 02} (2021)
  136}, [\href{http://arxiv.org/abs/2008.05274}{{\tt 2008.05274}}].

\bibitem{Balasubramanian:2020xqf}
V.~Balasubramanian, A.~Kar and T.~Ugajin, \emph{{Islands in de Sitter space}},
  \href{http://dx.doi.org/10.1007/JHEP02(2021)072}{\emph{JHEP} {\bf 02} (2021)
  072}, [\href{http://arxiv.org/abs/2008.05275}{{\tt 2008.05275}}].

\bibitem{Sybesma:2020fxg}
W.~Sybesma, \emph{{Pure de Sitter space and the island moving back in time}},
  \href{http://dx.doi.org/10.1088/1361-6382/abff9a}{\emph{Class. Quant. Grav.}
  {\bf 38} (2021) 145012}, [\href{http://arxiv.org/abs/2008.07994}{{\tt
  2008.07994}}].

\bibitem{Chen:2020hmv}
H.~Z. Chen, R.~C. Myers, D.~Neuenfeld, I.~A. Reyes and J.~Sandor,
  \emph{{Quantum Extremal Islands Made Easy, Part II: Black Holes on the
  Brane}}, \href{http://dx.doi.org/10.1007/JHEP12(2020)025}{\emph{JHEP} {\bf
  12} (2020) 025}, [\href{http://arxiv.org/abs/2010.00018}{{\tt 2010.00018}}].

\bibitem{Ling:2020laa}
Y.~Ling, Y.~Liu and Z.-Y. Xian, \emph{{Island in Charged Black Holes}},
  \href{http://arxiv.org/abs/2010.00037}{{\tt 2010.00037}}.

\bibitem{Marolf:2020rpm}
D.~Marolf and H.~Maxfield, \emph{{Observations of Hawking radiation: the Page
  curve and baby universes}},
  \href{http://dx.doi.org/10.1007/JHEP04(2021)272}{\emph{JHEP} {\bf 04} (2021)
  272}, [\href{http://arxiv.org/abs/2010.06602}{{\tt 2010.06602}}].

\bibitem{Hernandez:2020nem}
J.~Hernandez, R.~C. Myers and S.-M. Ruan, \emph{{Quantum extremal islands made
  easy. Part III. Complexity on the brane}},
  \href{http://dx.doi.org/10.1007/JHEP02(2021)173}{\emph{JHEP} {\bf 02} (2021)
  173}, [\href{http://arxiv.org/abs/2010.16398}{{\tt 2010.16398}}].

\bibitem{Matsuo:2020ypv}
Y.~Matsuo, \emph{{Islands and stretched horizon}},
  \href{http://arxiv.org/abs/2011.08814}{{\tt 2011.08814}}.

\bibitem{Akal:2020twv}
I.~Akal, Y.~Kusuki, N.~Shiba, T.~Takayanagi and Z.~Wei, \emph{{Entanglement
  Entropy in a Holographic Moving Mirror and the Page Curve}},
  \href{http://dx.doi.org/10.1103/PhysRevLett.126.061604}{\emph{Phys. Rev.
  Lett.} {\bf 126} (2021) 061604}, [\href{http://arxiv.org/abs/2011.12005}{{\tt
  2011.12005}}].

\bibitem{KumarBasak:2020ams}
J.~Kumar~Basak, D.~Basu, V.~Malvimat, H.~Parihar and G.~Sengupta,
  \emph{{Islands for Entanglement Negativity}},
  \href{http://arxiv.org/abs/2012.03983}{{\tt 2012.03983}}.

\bibitem{Caceres:2020jcn}
E.~Caceres, A.~Kundu, A.~K. Patra and S.~Shashi, \emph{{Warped Information and
  Entanglement Islands in AdS/WCFT}},
  \href{http://arxiv.org/abs/2012.05425}{{\tt 2012.05425}}.

\bibitem{Raju:2020smc}
S.~Raju, \emph{{Lessons from the Information Paradox}},
  \href{http://arxiv.org/abs/2012.05770}{{\tt 2012.05770}}.

\bibitem{Chu:2021gdb}
J.~Chu, F.~Deng and Y.~Zhou, \emph{{Page Curve from Defect Extremal Surface and
  Island in Higher Dimensions}},  \href{http://arxiv.org/abs/2105.09106}{{\tt
  2105.09106}}.

\bibitem{Bhattacharya:2020ymw}
A.~Bhattacharya, \emph{{Multipartite purification, multiboundary wormholes, and
  islands in $AdS_3/CFT_2$}},
  \href{http://dx.doi.org/10.1103/PhysRevD.102.046013}{\emph{Phys. Rev. D} {\bf
  102} (2020) 046013}, [\href{http://arxiv.org/abs/2003.11870}{{\tt
  2003.11870}}].

\bibitem{Bhattacharya:2020uun}
A.~Bhattacharya, A.~Chanda, S.~Maulik, C.~Northe and S.~Roy, \emph{{Topological
  shadows and complexity of islands in multiboundary wormholes}},
  \href{http://dx.doi.org/10.1007/JHEP02(2021)152}{\emph{JHEP} {\bf 02} (2021)
  152}, [\href{http://arxiv.org/abs/2010.04134}{{\tt 2010.04134}}].

\bibitem{Manu:2020tty}
A.~Manu, K.~Narayan and P.~Paul, \emph{{Cosmological singularities,
  entanglement and quantum extremal surfaces}},
  \href{http://dx.doi.org/10.1007/JHEP04(2021)200}{\emph{JHEP} {\bf 04} (2021)
  200}, [\href{http://arxiv.org/abs/2012.07351}{{\tt 2012.07351}}].

\bibitem{Wang:2021mqq}
X.~Wang, R.~Li and J.~Wang, \emph{{Page curves for a family of exactly solvable
  evaporating black holes}},
  \href{http://dx.doi.org/10.1103/PhysRevD.103.126026}{\emph{Phys. Rev. D} {\bf
  103} (2021) 126026}, [\href{http://arxiv.org/abs/2104.00224}{{\tt
  2104.00224}}].

\bibitem{Caceres:2021fuw}
E.~Caceres, A.~Kundu, A.~K. Patra and S.~Shashi, \emph{{Page Curves and Bath
  Deformations}},  \href{http://arxiv.org/abs/2107.00022}{{\tt 2107.00022}}.

\bibitem{Bhattacharya:2021jrn}
A.~Bhattacharya, A.~Bhattacharyya, P.~Nandy and A.~K. Patra, \emph{{Islands and
  complexity of eternal black hole and radiation subsystems for a doubly
  holographic model}},
  \href{http://dx.doi.org/10.1007/JHEP05(2021)135}{\emph{JHEP} {\bf 05} (2021)
  135}, [\href{http://arxiv.org/abs/2103.15852}{{\tt 2103.15852}}].

\bibitem{Ghosh:2021axl}
K.~Ghosh and C.~Krishnan, \emph{{Dirichlet Baths and the Not-so-Fine-Grained
  Page Curve}},  \href{http://arxiv.org/abs/2103.17253}{{\tt 2103.17253}}.

\bibitem{Geng:2020fxl}
H.~Geng, A.~Karch, C.~Perez-Pardavila, S.~Raju, L.~Randall, M.~Riojas et~al.,
  \emph{{Information Transfer with a Gravitating Bath}},
  \href{http://dx.doi.org/10.21468/SciPostPhys.10.5.103}{\emph{SciPost Phys.}
  {\bf 10} (2021) 103}, [\href{http://arxiv.org/abs/2012.04671}{{\tt
  2012.04671}}].

\bibitem{Geng:2021wcq}
H.~Geng, Y.~Nomura and H.-Y. Sun, \emph{{Information paradox and its resolution
  in de Sitter holography}},
  \href{http://dx.doi.org/10.1103/PhysRevD.103.126004}{\emph{Phys. Rev. D} {\bf
  103} (2021) 126004}, [\href{http://arxiv.org/abs/2103.07477}{{\tt
  2103.07477}}].

\bibitem{Geng:2021iyq}
H.~Geng, S.~L\"ust, R.~K. Mishra and D.~Wakeham, \emph{{Holographic BCFTs and
  Communicating Black Holes}},  \href{http://arxiv.org/abs/2104.07039}{{\tt
  2104.07039}}.

\bibitem{Geng:2021hlu}
H.~Geng, A.~Karch, C.~Perez-Pardavila, S.~Raju, L.~Randall, M.~Riojas et~al.,
  \emph{{Inconsistency of Islands in Theories with Long-Range Gravity}},
  \href{http://arxiv.org/abs/2107.03390}{{\tt 2107.03390}}.

\bibitem{Carroll:2009maa}
S.~M. Carroll, M.~C. Johnson and L.~Randall, \emph{{Extremal limits and black
  hole entropy}},
  \href{http://dx.doi.org/10.1088/1126-6708/2009/11/109}{\emph{JHEP} {\bf 11}
  (2009) 109}, [\href{http://arxiv.org/abs/0901.0931}{{\tt 0901.0931}}].

\bibitem{Sekino:2008he}
Y.~Sekino and L.~Susskind, \emph{{Fast Scramblers}},
  \href{http://dx.doi.org/10.1088/1126-6708/2008/10/065}{\emph{JHEP} {\bf 10}
  (2008) 065}, [\href{http://arxiv.org/abs/0808.2096}{{\tt 0808.2096}}].

\bibitem{Hayden_2007}
P.~Hayden and J.~Preskill, \emph{Black holes as mirrors: quantum information in
  random subsystems},
  \href{http://dx.doi.org/10.1088/1126-6708/2007/09/120}{\emph{Journal of High
  Energy Physics} {\bf 2007} (Sep, 2007) 120--120}.

\bibitem{Maldacena:2013xja}
J.~Maldacena and L.~Susskind, \emph{{Cool horizons for entangled black holes}},
  \href{http://dx.doi.org/10.1002/prop.201300020}{\emph{Fortsch. Phys.} {\bf
  61} (2013) 781--811}, [\href{http://arxiv.org/abs/1306.0533}{{\tt
  1306.0533}}].

\end{thebibliography}
\bibliographystyle{JHEP}

\providecommand{\href}[2]{#2}\begingroup\raggedright\endgroup

\end{document}